\documentclass[letterpaper]{./mandc2019}
\usepackage{graphicx}  
\usepackage{chngcntr}
\counterwithin{figure}{section}
\usepackage{amsmath}
\numberwithin{equation}{section}

\usepackage{amsfonts}
\usepackage{algorithm,algorithmic}
\usepackage{tabls,color}
\usepackage{cites}

\usepackage{epsf}
\usepackage{appendix}
\usepackage{ragged2e}
\usepackage{amsthm}
\usepackage{subcaption}
\usepackage[top=1in, bottom=1.in, left=1.in, right=1.in]{geometry}
\usepackage{enumitem}
\setlist[itemize]{leftmargin=*}
\usepackage{caption}
\usepackage[makeroom]{cancel}
\newcommand{\sigmat}{\sigma_\mathrm{t}}
\newcommand{\sigmas}{\sigma_\mathrm{s}}

\title{A low-rank method for time-dependent transport calculations}
\author{%
	\textbf{Zhuogang Peng$^1$, Ryan G. McClarren$^1$, and Martin Frank$^2$}
	\\
	$^1$University of Notre Dame, College of Engineering  \\
	Department of Aerospace and Mechanical Engineering \\ 
	Notre Dame, IN 46556 \\ 
	\\
	$^2$Karlsruhe Institute of Technology, Steinbuch Centre for Computing \\ 
	Karlsruhe, Germany\\ 
	\\
	\url{zpeng5@nd.edu}, \url{rmcclarr@nd.edu}, \url{martin.frank@kit.edu}
}
\newcommand{\authorHead}      
{First author}  
\newcommand{\shortTitle}      
{Paper Title }  
\begin{document}
	\maketitle
	\justify 
	
	\begin{abstract}
		Low-rank approximation is a technique to approximate a tensor or a matrix with a reduced rank to reduce the memory required and computational cost for simulation. Its broad applications include dimension reduction, signal processing, compression, and regression. In this work, a dynamical low-rank approximation method is developed for the time-dependent radiation transport equation in slab geometry. Using a finite volume discretization in space and Legendre polynomials in angle we construct a system that evolves on a low-rank manifold via an operator splitting approach. We demonstrate that the low-rank solution gives better accuracy than solving the full rank equations given the same amount of memory.  
	\end{abstract}
	\section{INTRODUCTION}
We consider the transport of neutral particles as described by the linear Boltzmann equation,
\begin{equation}\label{RadiativeTransfer}
\frac{1}{c}\frac{\partial \psi(z,\mu,t) }{\partial t} + \mu \frac{\partial  \psi(z,\mu,t)}{\partial z} + \sigmat \psi(z,\mu,t) = \frac{\sigmas }{2} 
\int_{-1}^1\psi(z,\mu',t) \,d\mu'+S.
\end{equation}
The total and isotropic scattering macroscopic cross-sections are denoted as $\sigma_t(z)$ and $\sigma_s(z)$, respectively, $c$ is the particle speed and $S(z,t)$ is a prescribed source. The angular flux $\psi(z,\mu, t)$ is a function of position $z$, time $t$, and the cosine of the polar angle $\mu \in [-1,1]$. We also write the scalar flux, $\phi(z,t)$, as the integral of the angular flux
\begin{equation}
\phi(z,t) = \int_{-1}^1\psi(z,\mu,t) \,d\mu.
\end{equation}


In slab geometry we have a two-dimensional phase space and in principle there is a general function space that can describe solutions to the transport problem. Intuitively, it is known that  many transport problems require only a subspace of this function space (called a manifold in mathematical parlance) to describe the transport. In other words, the solution is not any possible function of two variables, rather only a subset of functions. An example of this are problems in the diffusion limit: these problems require only a linear dependence on $\mu$.  One can also formulate problems where this manifold over which the solution depends evolves over time: a beam entering a scattering medium would be described by a delta-function in space and angle at time zero, but eventually relax to much smoother distribution that we could characterize using a simple basis expansion.

We desire to generalize this idea, and possibly automatically discover the manifold that describes the system evolution. We accomplish this task by expressing the solution to a transport problem as a basis expansion in space and angle, and using techniques to determine what subspace of those bases are needed to describe the solution and how that subspace evolves. We use the dynamical low-rank approximation (DLRA) of  Koch and Lubich to evolve time-dependent matrices  by tangent-space projection \cite{Koch2007}. DLRA has been extended  to tensors \cite{Koch2010} and further results can be found in \cite{Nonnenmacher2008}. DLRA has been used  to reduce the computational complexity of quantum propagation \cite{Kloss2017} by restricting the evolution to lower-rank amongst other work \cite{Jahnke2008,Boiveau2017,Einkemmer2018,Markovsky2008}. In this work, we apply DLRA to neutral particle transport.

Here we give a brief mathematical introduction of a robust and accurate projector-splitting method developed by Lubich \cite{Lubich2014} to perform the DLRA for matrix differential equations of the form
\[
\frac{\partial}{\partial t} A(t)  \equiv \dot{A}(t)= F(A(t)),
\]
for $A(t) \in \mathbb{R}^{m \times n}$.
DLRA seeks to find an approximating matrix $Y(t)$ of rank $r$ that minimizes the error in the Frobenius norm $||{Y}(t)-{A}(t)||_\mathrm{f}$. 
Then we note that rank $r$ matrices are a manifold, $\mathcal{M}_r$, of the space $\mathbb{R}^{m \times n}$. The solution to this minimization problem can be found using the singular value decomposition (SVD).  However, to use the SVD this way we would need to have the solution $A(t)$. 

We would prefer a way to evolve the solution on $\mathcal{M}_r$ directly. We reformulate the problem as minimizing the difference between the time derivative of the approximation and the solution 
$||\dot{Y}(t)-\dot{A}(t)||_\mathrm{f},$
where the derivative $\dot{Y}(t)$ is in the tangent space of $\mathcal{M}_r$. With a Galerkin condition the minimization problem is equivalent to an orthogonal projection. With the decomposition $Y(t) = U(t)S(t)V^T(t)$, this minimization problem can be solved using time splitting. 
%
%
	\section{NUMERICAL METHOD} 
	In this study we write the solution to Eq.~\eqref{RadiativeTransfer} as
	\begin{equation}\label{LRA1}
	\psi(z,\mu,t) \approx \sum_{i,j = 1}^r X_i(z,t)S_{ij}(t)W_j(\mu,t)
	\end{equation}
	as the best approximation with rank $r$ of the solution for the equation (\ref{RadiativeTransfer}), where we have written
	$X_i$ as an orthonormal basis for $z$ and $W_j$ as an orthonormal basis for $\mu$.
	We define the inner products \[\langle f,g\rangle _z = \int_{0}^{Z} f(z) g(z)\, dz, \qquad \langle f,g\rangle _\mu = \int_{-1}^1 f(\mu) g(\mu) \, d\mu.\] Due to orthonormality we also have  $\langle X_i, {X}_j \rangle_z = \langle W_i, {W}_j \rangle_\mu = \delta_{ij}.$ Then $
	\bar{X} = \{X_1, X_2, ..., X_r\}
	$
	and 
	$
	\bar{W} = \{W_1, W_2, ..., W_r\}
	$ are constructed as ansatz spaces.
    The expansion in Eq.~\eqref{LRA1} is not unique and we choose as gauge conditions  $\langle X_i, \dot{X}_j \rangle_z = 0$ and $\langle W_i, \dot{W}_j \rangle_\mu = 0$.
	We now define orthogonal projectors using the bases:
	\begin{equation}\label{Projection1}
	P_{\bar{X}} g = \sum_{i=1}^{r}X_i \langle X_i g \rangle_{z}
	\end{equation}
	\begin{equation}\label{Projection2}
	P_{\bar{W}} g = \sum_{j=1}^{r}W_j \langle W_j g \rangle_{\mu}
	\end{equation}

We apply the projectors to define a split of the original equations into three steps and each of these is solved for a short time step
	\begin{equation}\label{MainEq1}
	\partial_t \psi_1(z,\mu,t) = P_{\bar{W}}\left(-\mu\partial_z \psi_1(z,\mu,t)+\frac{\sigmas}{2}\int_{-1}^{1}\psi_1(z,\mu^{'},t)\,d\mu^{'}-\sigma_t \psi_1(z,\mu,t)+\frac{1}{2}S \right),
	\end{equation}
	\begin{equation}\label{MainEq2}
	\partial_t \psi_2(z,\mu,t) = -P_{\bar{X}}P_{\bar{W}}\left(-\mu\partial_z \psi_2(z,\mu,t)+\frac{\sigmas}{2}\int_{-1}^{1}\psi_2(z,\mu,t)d\mu^{'}-\sigmat \psi_2(z,\mu,t)+\frac{1}{2}S\right),
	\end{equation}
	\begin{equation}\label{MainEq3}
	\partial_t \psi_3(z,\mu,t) = P_{\bar{X}}\left(-\mu\partial_z \psi_3(z,\mu,t)+\frac{\sigmas}{2}\int_{-1}^{1}\psi_3(z,\mu^{'},t)d\mu^{'}-\sigmat \psi_3(z,\mu,t)+\frac{1}{2}S\right).
	\end{equation}
	The $\psi_2$ step uses $\psi_1$ as an initial condition, and the $\psi_3$ uses $\psi_2$ as an initial condition. 
	It can be shown that the above evolution is contained in the low rank manifold $\mathcal{M}_r$, if the initial value is in $\mathcal{M}_r$ \cite{Lubich2014} because the right-hand side of each step remains in the tangent space $\mathcal{T} \mathcal{M}_r$. 
	
	To make the splitting more concrete we write $\psi_1$ as
	\begin{equation}
\psi_1(z,\mu,t) = \sum_{j=1}^r K_j(z,t) W_j(\mu,t),
\end{equation}
where $K_j(z,t) = \sum^{r}_{i} X_{i}(z,t)S_{ij}(t)$. We plug this solution into Eq.~\eqref{MainEq1} and multiply by $W_\ell(\mu,t)$ and integrate over $\mu$ to get
\begin{equation}\label{eq:psi1}
\partial_t K_j  + K_j \cancelto{0}{\langle W_\ell \dot{W}_j \rangle_\mu} = - \sum_{j'=1}^r \langle \mu W_j W_{j'} \rangle_\mu \partial_z K_{j'} + \frac{\sigmas}{2} \sum_{j'=1}g^r \langle  W_j \rangle_\mu \langle W_{j'} \rangle_\mu K_{j'} - \sigmat K_j + \frac{\langle W_{j}\rangle_\mu}{2}S.
\end{equation}
Notice that there is no change in $W_j$ bases in this equation. Equation \eqref{eq:psi1} resembles the standard P$_N$ equations, a point we will return to later.  It is a system of advection problems in $z$. 

We can then factorize $K_j$ into $X_{i}^{(1)}$ and $S_{ij}^{(1)}$.  This is used to define an initial condition for $\psi_2 = \sum_{i,j=1}^r X_{i}^{(1)} S_{ij}^{(1)} W_j.$ Then, we can perform similar calculations on Eq.~\eqref{MainEq2} to get
\begin{multline}\label{eq:s_eq}
\partial_t S_{ij} = \sum_{kl}^{r}\langle \partial_z X_k X_i\rangle_z S_{kl} \langle \mu W_l W_j\rangle_{\mu} - \frac{1}{2}\sum_{kl}^{r}\langle \sigma_s X_k X_i\rangle_z S_{kl} \langle W_l \rangle_{\mu} \langle W_j \rangle_{\mu} + \\ \sigma_t S_{ij} - \frac{1}{2}\langle X_i  \rangle_{z}\langle W_j \rangle_{\mu}.
\end{multline}
We call this solution $S_{ij}^{(2)}$. Equation \eqref{eq:s_eq} is a set of $r^2$ ordinary differential equations.  The solution is used to create an initial condition for $\psi_3 = \sum_{i,j=1}^r X_{i}^{(1)} S_{ij}^{(2)} W_j.$ 

Writing $L_i = S_{ij}(t)W_{j}(t,\mu)$ we can multiply Eq.~\eqref{MainEq3} by a spatial basis function and integrate over space to get
	\begin{equation}\label{eq:L_eq}
	\partial_t L_{i} = -\mu \sum_{k}^{r} \langle \partial_z X_k X_i \rangle_{z} L_k + \frac{1}{2} \langle \sigma_s X_i\rangle_z \langle L_i \rangle_{\mu} - \langle \sigma_t X_i\rangle_z L_i + \frac{1}{2}\langle X_i S \rangle_{z},
	\end{equation}
	which evolves the solution in $\mu$ space. Upon factoring $L_i = S^{(3)}_{ij}(t)W_j^{(3)}(\mu,t)$, and write the solution as $\psi = \sum_{i,j,=1}^r X_i^{(1)}(x,t) S^{(3)}_{ij}(t)W_j^{(3)}(\mu,t).$
	
	\subsection{Discretization Details}
	The procedure outlined above of solving Eqs.~\eqref{eq:psi1}, \eqref{eq:s_eq}, and \eqref{eq:L_eq} in that order is accomplished by using a first-order explicit integration. The bases we use are based on a finite volume discretization in space with a constant mesh spacing $\Delta z$ and $m$ zones, and $n$ Legendre polynomials in angle. To make   orthonormal bases we define
\begin{equation}\label{Discretization1}
	X_i(t,z) = \sum_{k=1}^{m}Z_k(z)u_{ki}(t)
	\end{equation}
	\begin{equation}\label{Discretization2}
	W_j(t,\mu) = \sum_{l=1}^{n}P_l(\mu) v_{lj}(t)
	\end{equation}
	Noted that $Z_i(z) = \frac{1}{\sqrt{\Delta z}}$ with $z \in [z_{i-\frac{1}{2}},z_{i+\frac{1}{2}}]$ where $i$ is the cell number, $P_j(\mu) = \sqrt{\frac{2n-1}{2}}\bar{P}_{n-1}(\mu)$, where $\bar{P}_{n}(\mu)$ is the $n_{th}$ order Legendre polynomial, $u_{ki}$ and $v_{lj}$ are components of the time dependent matrix $U(t) \in \mathbb{R}^{m \times r}$ and  $V(t) \in \mathbb{R}^{n \times r}$. 
	After the first and last step in the split the matrices $U$ and $V$ found by a  QR decomposition to either $K_j$ or $L_i$. 
	
	The memory footprint required to compute the solution is the based on storing the matrices $U$, $V$ and $S$. Therefore, the memory required is \begin{equation}\label{Memory}
	\mathrm{memory} = 2(mr+r^2+nr),
	\end{equation} the factor 2 assumes that we need to store the previous step solution as well as the new step. The full solution to this problem without  splitting would require  a memory footprint of $2mn$. Therefore, for $r \ll m,n$ there will be large memory savings.
	
	In the solution procedure we needed to calculate $ \langle \mu W_j W_{j'} \rangle_\mu$.  Using our angular basis this term becomes.
	\begin{equation}\label{Term1}
	\begin{aligned}
	 \langle \mu W_j W_{j'} \rangle_\mu= & \left \langle \mu \sum_{i=1}^{n}P_i(\mu) v_{ij}(t)  \sum_{k=1}^{n}P_k(\mu) v_{kj'}(t) \right\rangle_{\mu} = \left\langle \sum_{i=1}^{n}\sum_{k=1}^{n}v_{ij}(t)\mu P_i(\mu) P_k(\mu) v_{kj'}(t)  \right\rangle_{\mu} \\
	= &  \sum_{i=1}^{n}\sum_{k=1}^{n}v_{il}(t) \langle \mu P_i(\mu) P_k(\mu) \rangle_{\mu} v_{kj}(t)
	\end{aligned}
	\end{equation}
	Note that $ \langle \mu P_i(\mu) P_k(\mu) \rangle_{\mu}$ forms a $n \times n$ matrix, $C$, that can be precomputed. Thus Eqs.~\eqref{Term1} requires $\mathcal{O}(n^2 r)$ operations, which is affordable because usually $n$ is not large and $C$ is sparse. Alternatively, we could calculate $ \langle \mu W_j W_{j'} \rangle_\mu$ on-the-fly by choosing $\mathcal{O}(n)$ quadrature points in angle, and it requires $\mathcal{O}(n r^2)$ operations for all the $r^2$ entries. 
	
	
	Additionally, using the standard upwinding technique for the spatial derivative terms leads to 
	\begin{align}\label{Term2}
	\partial_z K \langle \mu W^T W \rangle_{\mu} =& \frac{1}{\sqrt{\Delta z}}(K_{i+\frac{1}{2}} - K_{i-\frac{1}{2}}) \langle \mu W^T W \rangle_{\mu} \\
	&=\frac{1}{2\sqrt{\Delta z}}[(K_{i+1}-K_{i-1})V^T CV - (K_{i+1}-2K_{i}+K_{i-1})V^T \Sigma V],
	\end{align}
	where $\Sigma$ is a stabilization matrix that we take to be a diagonal matrix with the singular values of $C$. Other stabilization terms could be used, including Lax-Friedrichs where $V^T \Sigma V$ is replaced by a constant times an identity matrix.
	
	The spherical harmonic we used in the angular expansion can yield oscillatory or negative solutions. To address this issue we implemented angular filtering \cite{McClarren2010,radice2013new} which can significantly increase the performance of $P_n$ method in solving radiative transfer equation by removing the oscillations. We implemented the filter into our explicit solver and combined it with the low-rank approximation algorithm. The filtered equation adds anisotropic scattering. In this study we use a Lanczos filter.
		\subsection{Conservation}
	The low-rank algorithm we have described does not conserve the number of particles. This loss of conservation is a result of information lost in the algorithm when restricting the solution to low rank descriptions. We have addressed this by globally scaling the solution after each time step to correct for any particles lost.	This point is discussed in further detail in the conclusion section.
	\section{NUMERICAL RESULTS}
	\subsection{Plane source problem}
	\begin{figure}[h!] 
		\begin{subfigure}{1\textwidth}\caption{Low-rank solutions without a filter}
			\centering
			\includegraphics[scale=0.25]{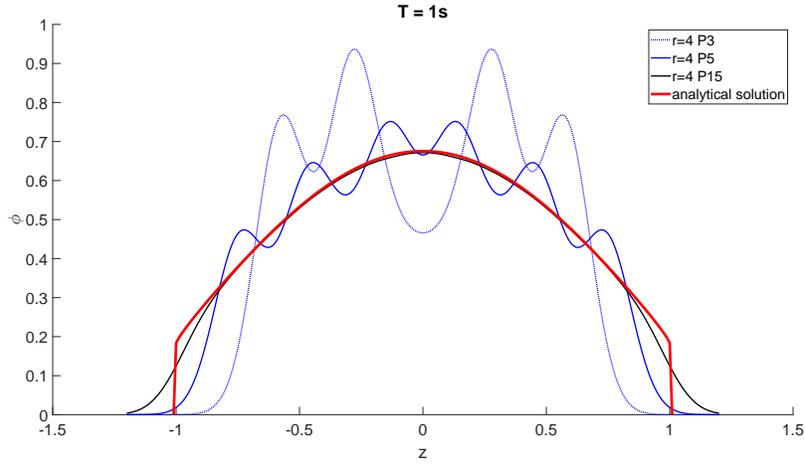}
		\end{subfigure}
		\begin{subfigure}{1\textwidth}\caption{Comparison of P$_7$ solutions of rank $4$ with and without a filter.}
			\centering
			\includegraphics[scale=0.25]{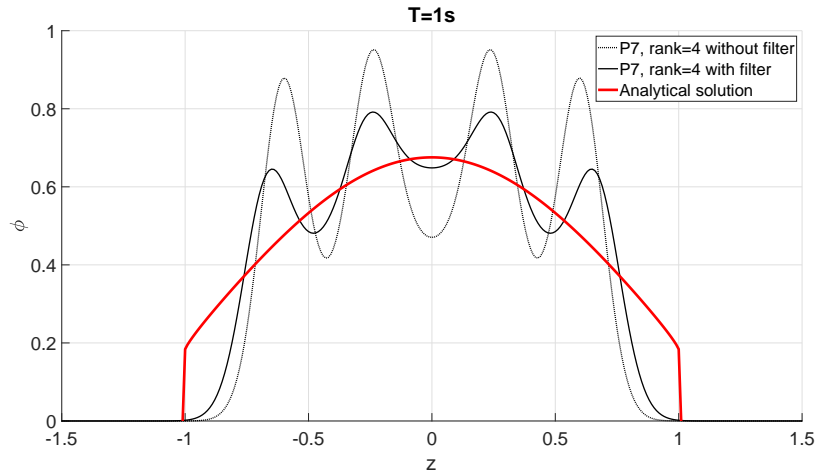}
		\end{subfigure}
		\caption{Solutions to the plane source problem using the low rank method compared to the analytic solution.}
		\label{PPresults}
	\end{figure}
	First, we solve the plane source problem of a delta function source in space and time in a purely scattering medium with $\sigmat = \sigmas = 1$;  the analytical benchmark solution was given by Ganapol \cite{Ganapol2008}. For this problem we fix the spatial resolution to be $\Delta z = 0.01$ (this corresponds to $m = 301$ for the $t=1$ solution and $m = 1201$ for the $t=5$ results), and vary the number of angular basis functions, $n$, and the rank $r$. When used, the filter strength is set to $50$.

	
	Figure \ref{PPresults} shows the solutions of varying rank and Legendre polynomial orders with and without a filter. We can see the low-rank solution using a P$_{15}$ basis  matches the analytic solution to the scale of the graph in the middle of the problem. We also observe that the low-rank solution can be improved by the filter: P$_7$ solutions of reduced rank improve when a filter is used. 
	
\begin{figure}[h!]
		\begin{subfigure}{1\textwidth}
			\centering
			\includegraphics[scale=0.25]{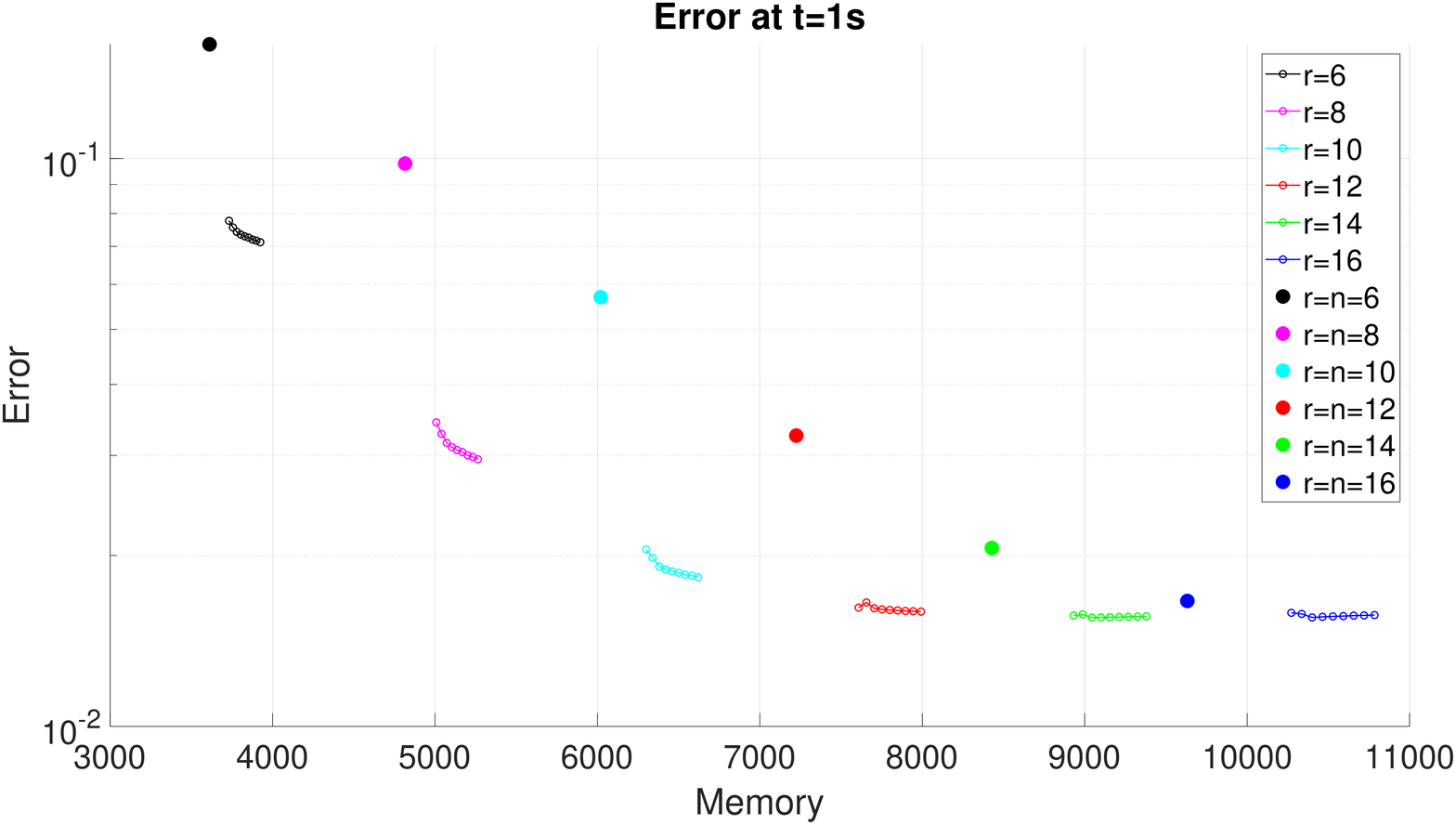}
		\end{subfigure}%
		
		\begin{subfigure}{1\textwidth}
			\centering
			\includegraphics[scale=0.25]{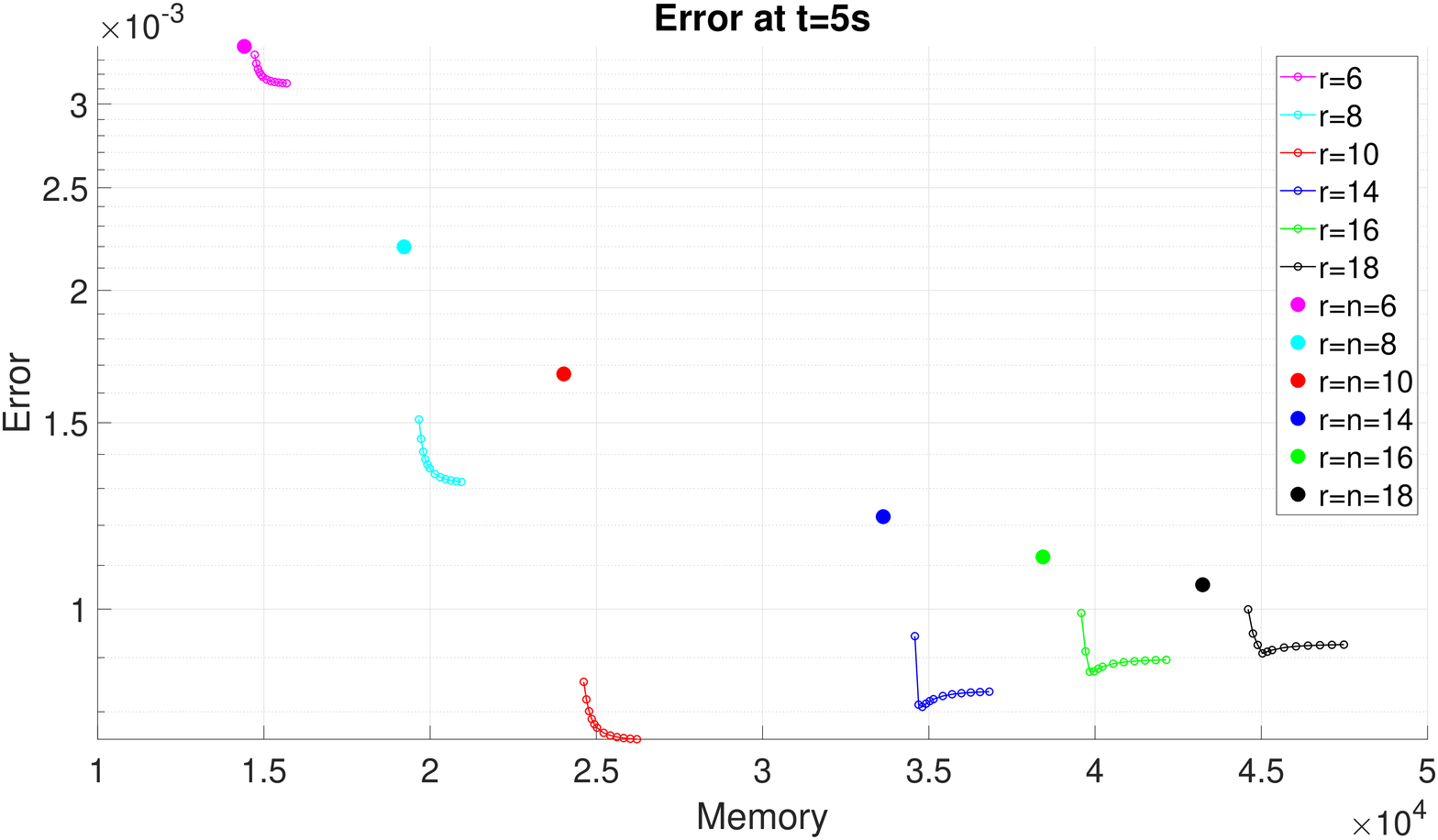}
		\end{subfigure}
		\caption{The comparison of errors on the plane source problem with different memory usage are shown. Each dotted line represents the error with a fixed rank that varies  the number of angular basis functions $n$. The bold dot denotes the full rank solution.}
		\label{PlanePulseErrors}	
	\end{figure}

	For a more quantitative comparison, the error of the numerical results with different $n$ and $r$ is shown in Figure \ref{PlanePulseErrors}. In this figure the colors for the dotted lines correspond to the rank used in a calculation and different values of the $n$, the number of angular basis functions, are corresponding dots. For each color the value of $n$ ranges from $r$ to $100$.  The large points are the value of the error using the standard full rank method with $r=n$. We can observe that the low-rank solution is more accurate than the full rank with the same memory usage. For example, the error of full rank solution $n=12$ using with a memory footprint of $8000$ is about $0.07$. With  less memory, the error can be reduced to $0.02$. We can also use $70\%$ of the memory to achieve the same accuracy. Increasing the resolution and rank will contribute to the accuracy of solutions. Given the way we performed this study with a fixed spatial mesh and time step and the conservation fix we used, we can see some error stagnation in the low-rank solution at $t=5$.  Other numerical experiments indicate that increasing the number of spatial zones can further decrease the error.
		
	\subsection{Reed's problem}
	\begin{figure} 
		\centering
		\includegraphics[scale=0.5]{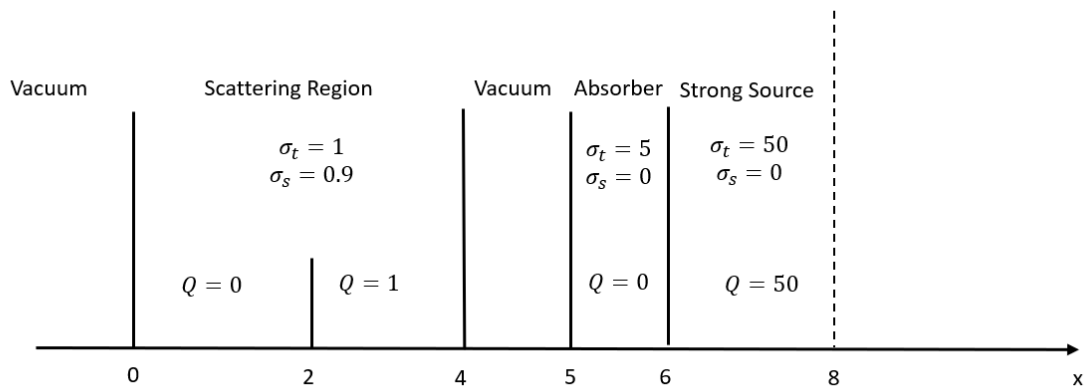} 
		\caption{The material layout in Reed’s problem.}   
		\label{Reeds}
	\end{figure}

\begin{figure} 
	\begin{subfigure}{1\textwidth}
		\centering
		\includegraphics[scale=0.25]{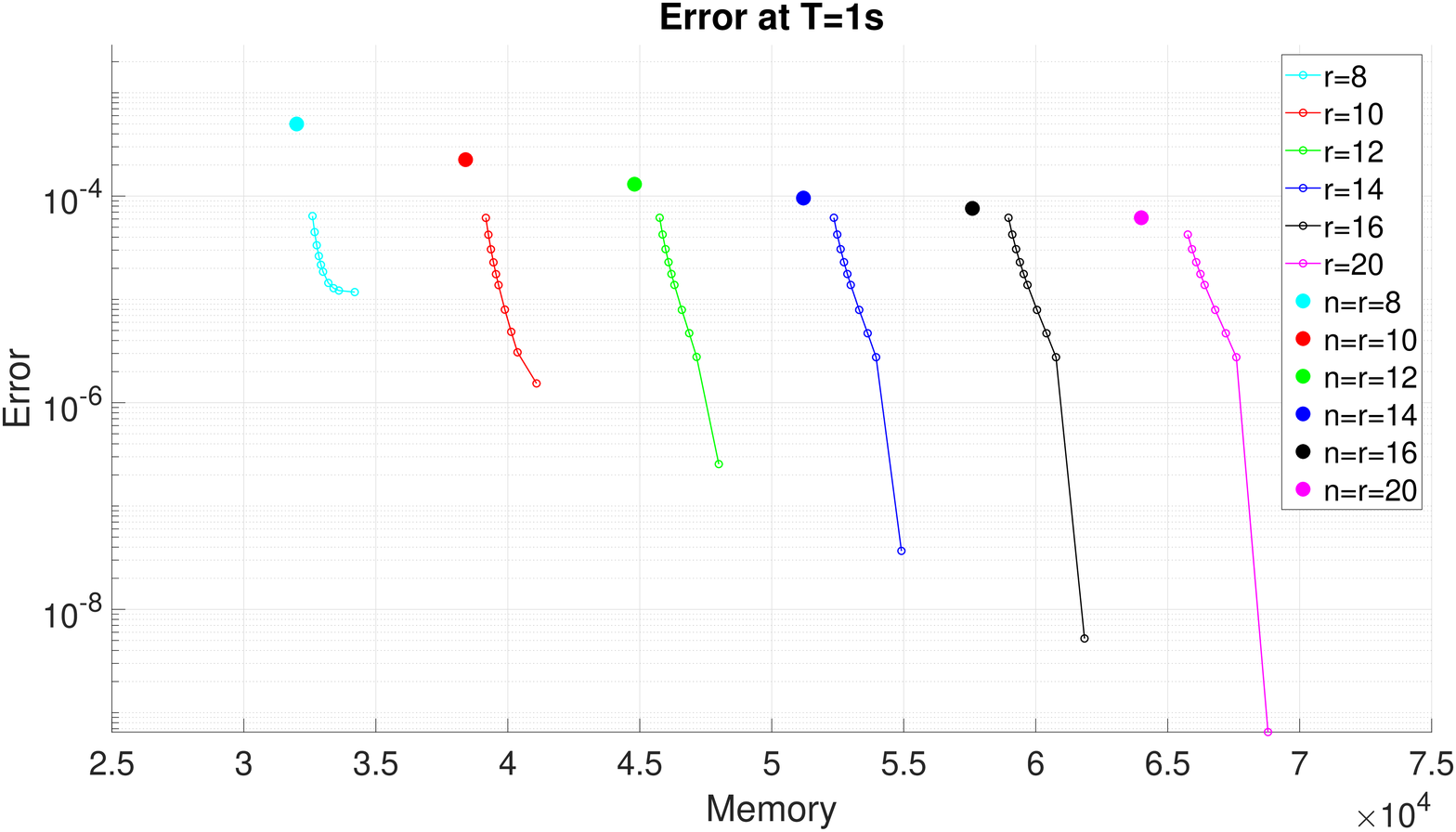}
	\end{subfigure}%
	
	\begin{subfigure}{1\textwidth}
		\centering
		\includegraphics[scale=0.25]{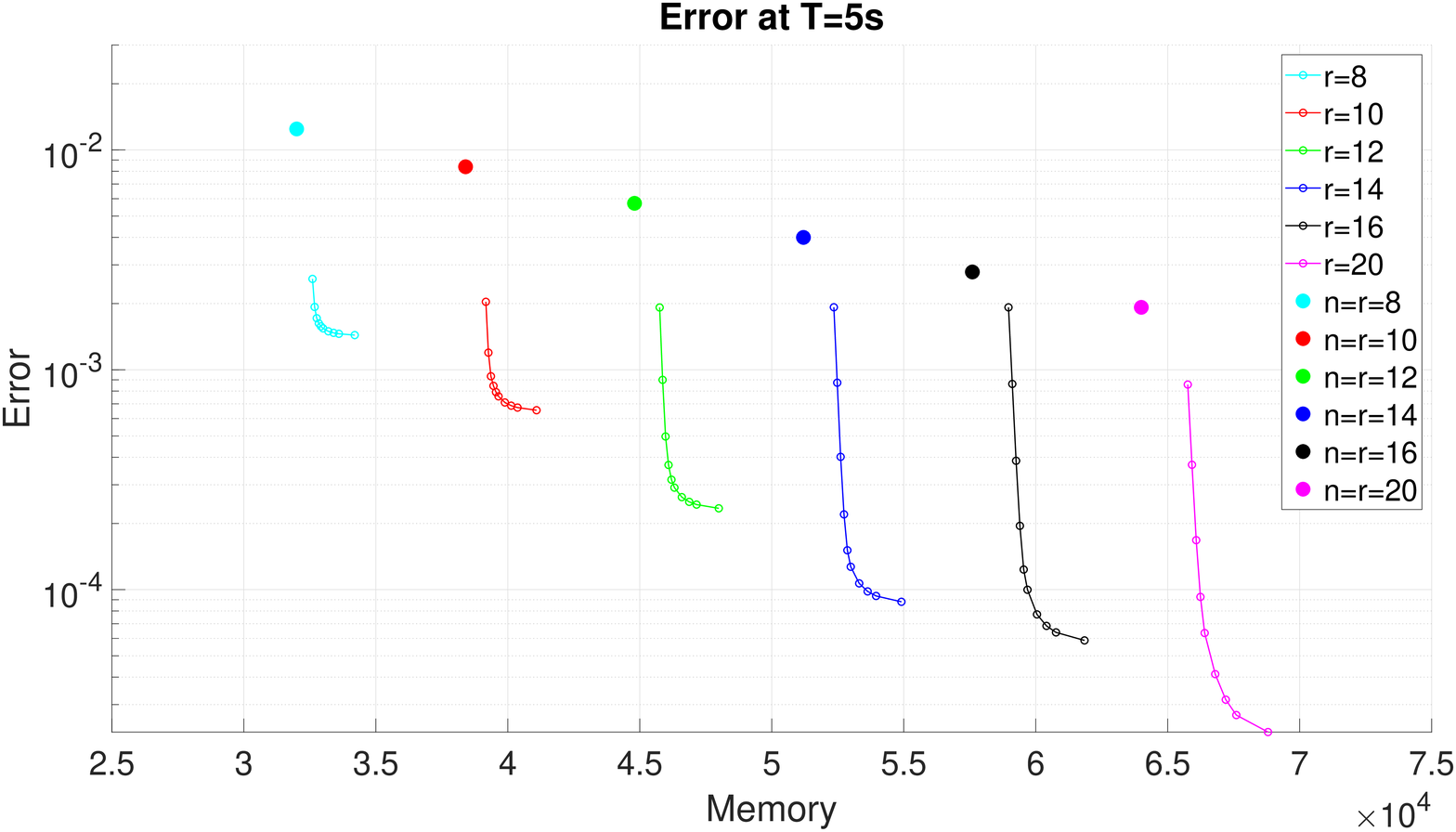}
	\end{subfigure}
	\caption{The comparison of errors for Reed's problem with different memory usage are shown. Each dotted line represents the error with a fixed rank that varies  the number of angular basis functions $n$. The bold dot denotes the full rank solution.}   
	\label{Reedserror}
    \end{figure}

	The second problem is Reed's problem \cite{Reed,Mcclarren2007,Zheng:2016ei}, which is a multi-material problem, and its set-up is detailed in Fig.~\ref{Reeds}. 
	Because Reed's problem does not have an analytical solution, a numerical result with high resolution and full rank, where $\Delta z = 0.01$ ($m=1600$), P$_{99}$ ($n= 100$) and $CFL = \Delta t / \Delta x = 0.1$, is set as a benchmark for memory analysis. It can be observed in Figure \ref{Reedserror} that the low rank solutions (solid lines with small dots) can give solutions with comparable errors to the full rank solutions (large dots) with much larger memory. For example the rank $8$ solutions obtain a solution error better than the full rank $P_{19}$ solution with less memory.
	
	
	\section{CONCLUSIONS}
	
	We have developed a practical algorithm to find the low-rank solution of the slab geometry transport equation using  explicit time integration. The method is based on projecting the equation to low-rank manifolds and numerically integrating in three steps. The numerical simulations show that on several test problems the memory savings of the low-rank method can be on the order of a factor of 2-3. Given that these are only slab geometry problems we  expect even larger memory savings on 2- and 3-D problems due to their larger size. Exploring this is ongoing work. Furthermore, we will be investigating other means for correcting the loss of conservation in the method, including posing the problem as a high-order/low-order problem.
	
	\setlength{\baselineskip}{12pt}
	\bibliographystyle{mandc}
    
	\bibliography{mandc_references}
\end{document}